# Modeling and Analysis of Network Dynamics in Complex Communication Networks Using Social Network Methods

Bisma S. Khan, Muaz A. Niazi*

*Abstract*—Modern communication networks are inherently complex in nature. First of all, they have a large number of heterogeneous components. Secondly, their connectivity is extremely dynamic. Nodes can come and go, links can be removed and added over time. Traditional modeling and simulation techniques amalgamate or ignore such dynamics and therefore, are unable to represent them. Complex communication networks are therefore better modeled as mathematical structures called graphs. Modeling as graphs allows for the application of complex techniques such as various network based analysis techniques. While this is a very important and much needed skill for communication networks researchers and engineers, to the best of our knowledge, currently there is no resource describing these details. In this paper, we give a concise but comprehensive review of modeling complex communication networks as graphs. We also show how to apply complex social network analysis on these models besides a demonstration of formal modeling network dynamics.

*Corresponding author: muaz.niazi@gmail.com

## I. INTRODUCTION

Arguably, the most suitable representation of communication networks is in the form of graphs [1] or adaptive networks [2]. An alternative modeling technique is using differential equations. However, differential equations do not have the capabilities to properly model the intricacies of complex networks. When we convert a network into a differential equation model, all devices and systems are reduced to numbers/variables. Whereas, in graphs, all devices and links are preserved. Same is the case for networks represented only as samples of parameters. While these samples and tables give statistical results, they do not give any internal details of the system. The interactions in complex communication networks can be modeled as links and machines/devices/sensors/servers as nodes. The benefits of modeling networks as graphs include the reuse and adaptation of well-established techniques from this domain.

A complex network is a very well established and dynamic field. They provide an adequate theoretical framework for observing the structure and emerging behavior of naturally occurring networks. However, traditional approaches used in complex networks are not typically concerned with the structure and dynamics of real-world networks. We can apply computational techniques of complex networks on communication networks. Quantitative as well as visual information about the topological organization of the communication networks can be obtained through complex networks.

In complex communication networks, a huge number of changes occur in the network over time. Individual elements in a communication network give rise to the unpredictable and unprecedented behaviors. For example, at the physical layer, web server machines could be on and off and switches on the data link layer may be enabled and disabled. Likewise, at the network layer, routers could be on or off and connected or disconnected. The same way, at the session layer, the endpoint connections may establish or terminate. In like manner, at the presentation layer, the link between the HTTP daemon and web server may be up or down. Similarly, at the application layer, application programs may be opened or closed. This behavior cannot be matched in simple simulations based on few snapshots of the network. Certain features of the communication networks can be modeled, but it is very difficult to model entire communication network and its emergence because the exact nature of network dynamics cannot be modeled.

While the modeling of networks is a very important domain, typically, communication engineers and researchers are not conversant with the techniques and terminologies associated with modeling and analyzing communication systems using complex networks. Although the domain of complex networks is still growing rapidly and articles are publishing daily, a substantial literature has already accumulated. This drives the necessity to summarize the necessary background to understand how to model, analyze and simulate communication systems as networks. In this letter, we present an accessible introduction to the basic tools, techniques, and terminologies from the domain of complex/social network for communication engineers, network designers and practitioners allowing for using modeling to develop better and more scalable real-world large-scale networks.

The remaining paper is outlined as: Section II demonstrates modeling as networks. Section III describes where to go from here.

## II. MODELING AS NETWORKS

### A. Network Basics

Here we present formal representation of the network. Mathematically, a complex communication network can be conceptualized as a graph *G (N, L)*, where *N* is a collection of nodes (vertices) and *L* is a collection of links (lines). Each unit (agent or object) in the network is thought of as nodes (vertices) and the interaction or the relation between the pair of entities is thought of as links. A network *g(v, e)* is termed as a *subnetwork* if *v* is the subset of *N* and *e* is the subset of *L* [3]. An undirected link is termed as *edge* and a directed link is termed as an *arc*. The order of *N* and *L* is denoted by *n* and *m* respectively.

Let *A* is the *adjacency matrix*, where $A_{ij}$ is the entry in *A*. Two nodes *i* and *j* are adjacent if they are directly connected to each other. The entry $A_{ij} = 1$, if *i* and *j* are immediate neighbors, otherwise entry $A_{ij} = 0$. A *weighted network* has weight *w* attached to the edges, where *w* is a real number [4].

#### 1) Basic Definitions

Here, we briefly review some of the important concepts of networks. *Path* in a network is the series of consecutive edges [3]. The path length is the total number of links in that path. The *cycle* is a path in which starting and ending nodes are the same [4]. A *loop* in a network is a link that connects a node to itself [4]. A network is said to be disconnected if there is no path between subnetworks [4]. Conversely, a network is called to be connected if there exists a path between every pair of vertices [4]. Figure 3(a) depicts a connected network, whereas Fig. 3(h) displays a disconnected network. A *maximal subnetwork* cannot be extended while preserving its structural characteristic [4]. A *maximal connected network* is maximal in terms of connectedness [4]. A complete subnetwork containing two nodes is called a *dyad*, whereas a complete subnetwork containing three nodes is called a *complete triad [4]*. A *maximal complete network* is maximal in terms of completeness [5].

### B. Complex Network Models

To date, four basic types of complex network models are typically considered. They are as follows:
  i.   Regular Graphs [6]
  ii.  Erdős-Rényi (ER) networks [7]
  iii. Barabási-Albert (BA) Model [8]
  iv.  Watts-Strogatz (WS) Model [9]

In Regular Graphs, every vertex has the same degree, such as lattice and crystals. The ER models are completely random and are formed based on probability *p*. BA Models generate the graph with scale-free property. Some vertices have many connections with other vertices, whereas others have fewer connections with other vertices. They follow power-law degree distribution. Examples of these networks include co-authorship, word frequency, WWW, etc. WS model generates random graphs with the small-world property, where every node is at a short distance from every other node in terms of hops, for example, the neural network of C. Elegan, metabolic networks, actors network, etc. Real world networks often exhibit many of the properties of these networks.

## C. Complex Network Analysis Techniques

This subsection presents a brief overview of complex network analysis techniques. Certain aspects of the communication networks can be modeled by using simulation techniques. However, complex network modeling techniques may be more suitable for modeling the underlying dynamics and inherent complexity in the real-world communication networks. But, it is not applied previously to model and analyze network dynamics in communication networks.

### 1) Cohesion

This subsection focuses on social cohesion. Intuitively, cohesion signifies that the social network is composed of many ties. Cohesion is related to the structural concept of connectedness and density. The number of connections incident to a vertex is called a *degree*. Network *density* is characterized as the ratio of the actual edges in the network to the potential edges [3].

#### a) Cohesive Subgroups

*Cohesive subgroups* are the groups of actors who stick together (united) through direct, strong, and frequent ties. Most of the techniques for the identification of subgroups are based on the connectedness and density. Four of these are demonstrated below.

- *Components:* A network can comprise of two or more disconnected subnetworks. A subnetwork of this kind will be a component if it is a maximal connected network. Each node can belong to at most one component [4, 10]. Figures 3(h), 2(i), and 2(b) contain 2, 4, and 3 components respectively.
- A *k-core* is a maximal subnetwork where each node is incident with at least *k* other nodes within the subnetwork. *K*-cores are nested, members of *k*-cores will also be the members of (*k-1*)-cores and so on. A node may belong to multiple cores. A *k*-core may not be a connected subnetwork, members of one core may belong to several components [4, 10]. The network shown in Fig. 2(a) contains 2-cores.
- A *clique* is a maximally complete subnetwork containing at least three nodes. Unfortunately, it is hard to find cliques from large scale networks [4, 10]. Figure 2(a) contains two cliques: *V2, V6, V9* and *V2, V7, V9*.
- *Communities* in a network are the dense groups of vertices, which are highly connected to each other inside group than to the rest of the vertices in the network. Communities can be both overlapping or disjoint [3]. Figure 2(a) contains two communities: *V1, V2, V6, V7, V8, V9* and *V3, V4, V5, V8*.

*b)    Sentiments and Friendship*

In social networks, relations can be either positive of negative, for example, liking vs disliking and friendship vs hostility. Such relations are called *affective relations* and the study of such relations is called *sociometry*. "A *signed graph* is a graph in which each line carries either a positive or a negative sign" [4]. A sign is assigned to a given link based on the positive or negative attitude of the source of the link to the destination. A signed graph is called *balanced graph* if all positive ties occur inside the group and all negative ties occur between the groups [4].

*c)    Affiliations*

An *affiliation* network is a two-mode network in which the members are connected to one another via co-participation in some type of event or co-membership in a group. Affiliation networks contain two or more sets of nodes; affiliations join nodes from different sets only. Usually, there are two sets called *actors* (people) and *events* (organizations) [4, 10].

*2) Brokerage*

Social networks as structures transmit information between actors. Information diffuse in the network through central nodes, which have important positions in the network. The number and intensity of an actor's ties are known as *social capital* [4]. It is important to know central actors in the communication network.

*a)    Centrality*

Centrality is the most important topological characteristic of a node which influences the network dynamics. It basically measures the topological significance of the individual nodes in the network [4]. Some of the commonly used centralities are described below:

- *Degree Centrality (DC):* It is the number of the immediate neighbors (connections) incident with a vertex. Degree centrality considers every neighbor as one "centrality point" [4, 11]. However, all neighbors are not alike; the importance of a node may increase if it is connected to the important neighbors.
- *Eigenvector Centrality (EC):* The idea is to measure the influence (centrality) of a node in terms of the sum of the centralities of its neighbors. The EC of a given node increases if it has many or important neighbors. The EC is defined as "the principal or dominant eigenvector of the adjacency matrix *A* representing the connected subgraph or component of a network" [11, 12].
- *Closeness centrality [13]:* It is based on the concept that how close an individual node is to the other nodes in the network. The CC of a vertex *v* is the inverse of the average geodesic distances between *v* to the rest of vertices in the network [4, 11].

- *Betweenness centrality (BC):* It determines the extent to which a vertex is included in the geodesics of the pairs of the all other vertices [4, 11].
- *Eccentricity centrality (ECC):* It determines the maximum geodesic distance between a vertex and all other vertices [14].

Table I lists the centrality scores of Fig. 2(a) computed by three different tools: Mathematica, R, and Visone.

TABLE I.

THE BETWEENNESS CENTRALITY, CLOSENESS CENTRALITY, DEGREE CENTRALITY, ECCENTRICITY, AND EIGENVECTOR CENTRALITY SCORES OF THE NODES OF THE RANDOM NETWORK SHOWN IN FIG. 3A. THREE DIFFERENT TOOLS MATHEMATICA(M), R, AND VISONE (V) ARE USED.

| Id | Betweenness | | | Closeness | | | Degree | | | Eccentricity | | | Eigenvector | | |
|---|---|---|---|---|---|---|---|---|---|---|---|---|---|---|---|
|  | M | R | V | M | R | V | M | R | V | M | R | V | M | R | V (%) |
| V1 | 0 | 0 | 0 | 0.428 | 0.428 | 0.429 | 1 | 1 | 0.111 | 0.25 | 4 | 2.25 | 0.063 | 0.316 | 6.399 |
| V2 | 7.5 | 7.5 | 0.208 | 0.6 | 0.6 | 0.6 | 4 | 4 | 0.444 | 0.333 | 3 | 3 | 0.1691 | 0.8737 | 16.912 |
| V3 | 9 | 9 | 0.25 | 0.529 | 0.529 | 0.529 | 3 | 3 | 0.333 | 0.333 | 3 | 3 | 0.096 | 0.478 | 9.663 |
| V4 | 0 | 0 | 0 | 0.36 | 0.36 | 0.36 | 1 | 1 | 0.111 | 0.25 | 4 | 2.25 | 0.030 | 0.151 | 3.063 |
| V5 | 11 | 11 | 0.306 | 0.6 | 0.6 | 0.6 | 3 | 3 | 0.333 | 0.5 | 2 | 4.5 | 0.105 | 0.520 | 10.519 |
| V6 | 0 | 0 | 0 | 0.5 | 0.5 | 0.5 | 2 | 2 | 0.222 | 0.333 | 3 | 3 | 0.117 | 0.582 | 11.757 |
| V7 | 0 | 0 | 0 | 0.5 | 0.5 | 0.5 | 2 | 2 | 0.222 | 0.333 | 3 | 3 | 0.117 | 0.582 | 11.757 |
| V8 | 0 | 0 | 0 | 0.391 | 0.391 | 0.391 | 1 | 1 | 0.111 | 0.333 | 3 | 3 | 0.033 | 0.164 | 3.334 |
| V9 | 20.5 | 20.5 | 0.569 | 0.692 | 0.692 | 0.692 | 6 | 6 | 0.667 | 0.333 | 3 | 3 | 0.202 | 1 | 20.197 |
| V10 | 0 | 0 | 0 | 0.428 | 0.428 | 0.429 | 1 | 1 | 0.111 | 0.25 | 4 | 2.25 | 0.063 | 0.316 | 6.399 |

b) *Bridges and Bi-Components*

Actors on crucial positions in the social networks have control over strategic information diffusion.

- A *bridge* is a link in the network whose removal results in an increased number of components. A bridge may act as a bottleneck in the network, which prohibit the flow of information [4, 10]. In Fig 2(f), link *V5-V9* is one of the bridges.
- A *cut-vertex* (or articulation point) is a node in the network whose removal results in increased number of the components [4]. In Fig. 3(g), V9 is one of the articulation points.
- A *bi-component* is a component which contains at least *3* nodes and does not have an articulation point [4]. In Fig. 2(g), subnetwork (*V1, V2, V6, V9*) is a bi-component.

c) *Ego*

The ego network tells us about individual nodes instead of the entire network. The ego network contains a focal vertex (ego), ego's neighbors (alters), and ties among alters. A triad is comprised of an ego, an alter, a third actor, and ties among them. Ego plays a brokerage role between pairs of alters which are not directly connected. A broker may be at a powerful and advantageous position in an incomplete triad. The ego network contains structural holes, which may be exploited. A *structural hole* is the absence of connection (tie) between an alter and the third actor [4].

*3) Brokerage Roles*

Here we investigate five possible positions of the broker in the directed triad.

1. In *coordinator role*, the broker is at the intermediary position as shown in Fig. 1(a). Broker and other two actors are the members of the same group [4].
2. In *itinerant role*, the broker is at the intermediary position and is brokering between other two actors as shown in Fig. 1(b). The other two actors belong to the same group, but broker itself is not a member of that group [4].
3. In *representative role*, the broker and one actor belong to the same group, while another actor belongs to a different group as shown in Fig. 1(c). The broker is positioned at the boundary of its group and it controls the flow of information from its group [4].
4. In *gatekeeper role*, the broker and one actor belong to one group, while another actor belongs to a different group as shown in Fig. 1(d). The broker is positioned at the boundary of its group and it controls the flow of the information to its group [4].
5. **In *liaison role*, the broker is in the intermediary position between two groups as shown in Fig. 1(e). Each group contains one member; broker does not belong to either group [4].**

*4) Diffusion*

Diffusion demonstrates the dissemination of events (information, disease, rumor, or virus, etc.) all over a network. The vertex *V9* in Fig. 3(a) can be a good source of diffusion. Social networks are an active source of information dissemination. On one hand, useful information can be promoted and propagated efficiently and effectively via social networks. At the other hand, malicious information such as virus and rumor can also propagate uncontrollably in social networks.

*5) Prestige*

Prestige measures the structural prominence of an individual actor in the directed network. In order to quantify the prominence of an actor, it is required to examine all its direct and indirect ties. A prestigious actor receives many ties from others, but sends few ties [4].

- *Indegree (popularity)* of an actor is the number of links it receives from the other actors. Popularity is the measure of structural prestige. However, indegree prestige does not consider indirect ties, it only considers direct ties. Therefore, it disregards the overall network structure [4].
- *Input (Influence Domain)* of an actor is the number of all other actors who are directly or indirectly linked

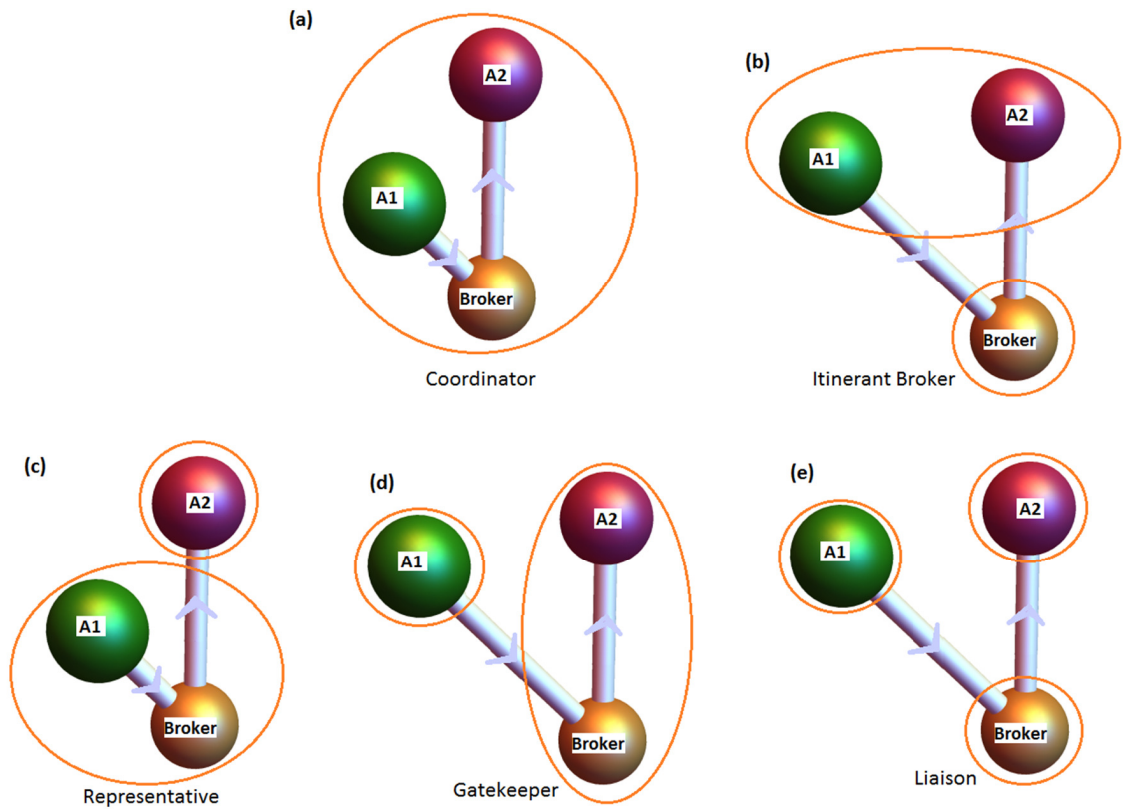

Figure 1. Five possible positions of the broker in a directed triad. a) Coordinator; b) Itinerant Broker; c) Representative; d) Liaison; e) Gatekeeper. This figure is adapted from [2]. The Mathematica is used to draw network models, contours around the nodes are drawn manually.

to this actor. However, influence domain is also not a perfect prestige measure. In a well-connected network, it may count all or all most all actors, so it does not differentiate well between actors [4].

- *The proximity prestige* of an actor is the fraction of all other actors in its influence domain divided by the average distance from all actors in its influence domain. It limits the influence domain to the immediate neighbors or to the neighbors of the neighbors [4].

*6) Ranking*

The social ranking may be formal, informal, or a combination of both. In the *formal ranking*, the insignia is assigned to individuals to show their particular rank, group, or organization. An example of formal ranking is an army. Conversely, the *informal ranking* is neither expressed by official symbols or marks nor is written down. It manifests itself by the behavior and opinion of people, such as deference and dominance [4]. Social network analysis techniques are used to investigate informal ranking of people.

The structural concept of ranking is based on balance theory. According to the *balance theory*, people tend to like each other inside the cluster, however, they do not like the rest of the members in other clusters. Balance theory can be rephrased for dyads in directed networks [4]. The *dyad* is a subnetwork in diagraph consisting of a pair of nodes and link between them.

There are three isomorphism classes of dyads [4].

1. *Null Dyad* is an edgeless subgraph. Two clusters may be separated by null dyad.
2. *Mutual Dyads* are complete dyads with two arcs going in either direction. As mutual dyads represent reciprocal links, so it is believed that both nodes in the mutual dyad belong to the same rank.
3. *Asymmetric Dyad* has one link going in one direction. It is considered that the receiving node in the asymmetric dyad, has a higher rank.

*D. From Systems to Networks: Cognitive Agent-based Computing*

In several previous special issues of journals, articles such as [15], and related volumes such as [16], we have presented the concepts related to modeling and simulation. Complex Adaptive COmmunicatiOn Networks and environmentS (CACOONS) is a general term for large-scale communication networks which exhibit some or all of complexity features of Complex Adaptive Systems [16]. Cognitive agent based computing (CABC) framework [16] is a recently proposed unified framework which allows modeling and analysis of artificial and natural Complex Adaptive Systems. The CABC combines Agent-Based Modeling and Complex Networks Based Models to develop cognition of the CAS. Here we present a case study.

*1) Modeling Network Dynamics: A Practical Case Study*

We observe, what changes can occur in a communication network over time. Let *G* be a random, undirected, and unweighted network. The change in the network over time can be one of these types: +*N*, -*N*, ±*N*, +*L*, -*L*, ±*L*, where +*N* represents addition of node(s) in the network, -*N* represents removal of node(s) from the network, ±*N* represents addition and subtraction of nodes in the network, +*L* represents addition of link(s) in the network, -*L* represents removal of link(s) in the network, and ±*L* represents addition and subtraction of links in the network. There are *15* possible changes in the network as demonstrated in Table II.

TABLE II.

THE NETWORK DYNAMICS IN COMMUNICATION NETWORKS AS A RESULT OF EVOLUTION OR SHRINKING OF THE NETWORK OVER TIME BY THE ADDITION OR REMOVAL OF VERTICES (N) OR EDGES (L).

| Model | Nodes | Links | Description |
|---|---|---|---|
| 1 | N | - | A random network |
| 2 | +N | - | Addition of node(s) in the network |
| 3 | -N | - | Removal of node(s) in the network |
| 4 | ±N | - | Addition and removal of nodes in the network |
| 5 | - | +L | Addition of link(s) in the network |
| 6 | - | -L | Removal of link(s) in the network |
| 7 | - | ±L | Addition and removal of links in the network |
| 8 | +N | +L | Addition of node(s) and link(s) in the network |
| 9 | +N | -L | Addition of node(s) and removal of link(s) in the network |
| 10 | +N | ±L | Addition of node(s) and addition and removal of link(s) in the network |
| 11 | -N | +L | Removal of node(s) and addition of link(s) in the network |
| 12 | -N | -L | Removal of node(s) and removal of link(s) in the network |
| 13 | -N | ±L | Removal of node(s) and addition of link(s) and removal of link(s) in the network |
| 14 | ±N | +L | Addition and removal of nodes and addition of link(s) in the network |
| 15 | ±N | -L | Addition and removal of nodes and removal of link(s) in the network |
| 16 | ±N | ±L | Addition and removal of nodes and addition and removal of links in the network |

Here, we address how the structure of the network changes over addition or deletion of the node(s) or link(s). We have used Wolfram Mathematica, a modern technical computing development platform for modeling. New nodes and links emerged in the network are marked by the red dotted circles and red arrows respectively. Whereas, nodes and links removed from the network are highlighted by red rectangles and red dotted lines respectively. Figures 2(a) and 3(a) display a random undirected and unweighted graph *G*. It contains *n=10* and *m=12*.

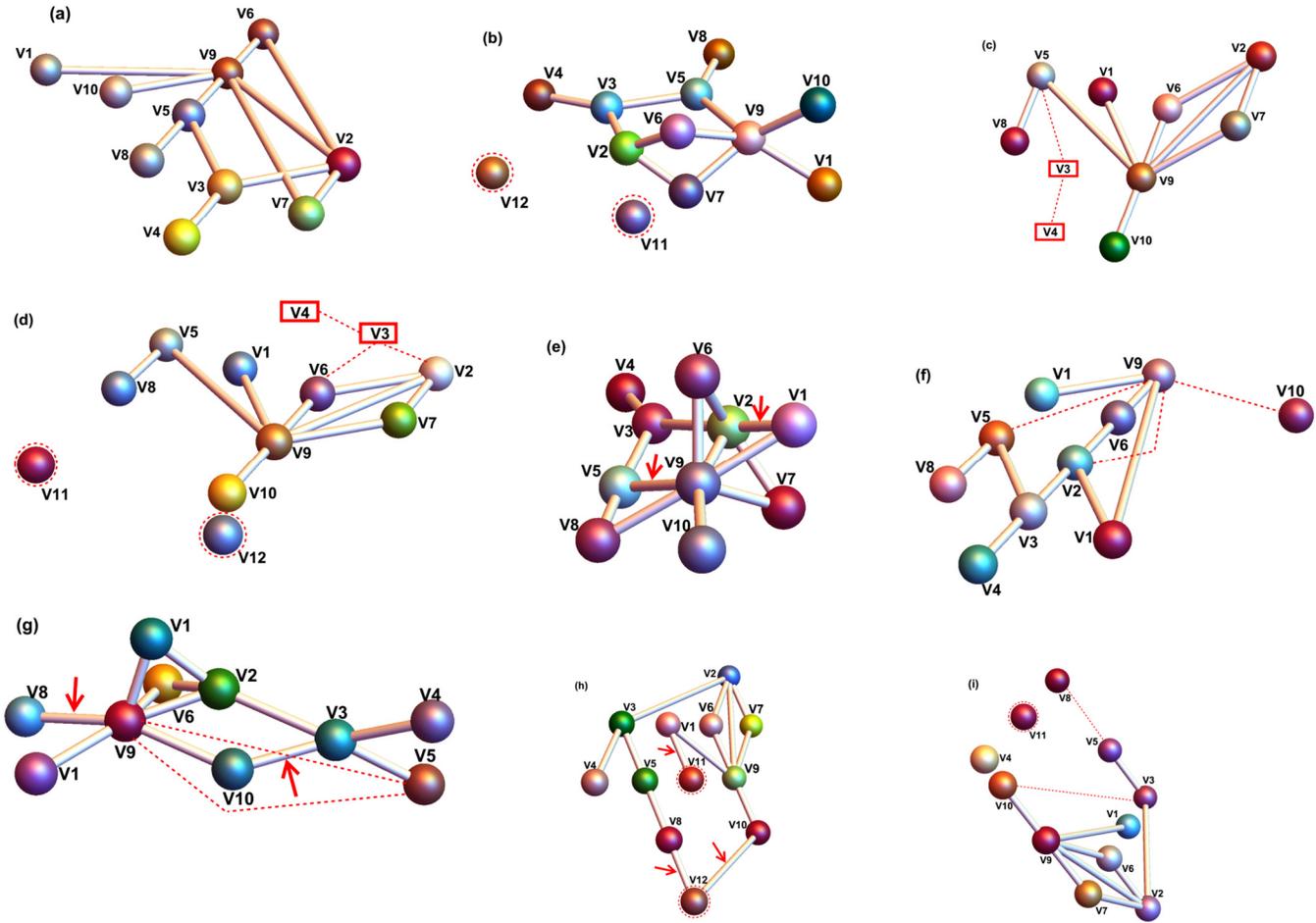

**Figure 2.** The dotted circles and rectangle nodes represent added and deleted nodes respectively. Red arrows and dotted lines represent the addition and deletion of links in the network respectively. a) The random network *G*, containing *N=10* and *L=12;* b) The network obtained after addition of new nodes as well as links and removal of links in *G*; c) Network obtained after removal of nodes and addition of new links in *G*; d) Network obtained after removal of existing links and nodes in *G*; e) The network obtained after addition of new links as well as the new nodes and removal of existing links in *G*; f) Network obtained after addition of new nodes as well as the links and removal of some existing nodes in *G*; g) Network obtained after addition of new nodes and removal of some existing nodes as well as links in *G*; h) Network obtained after addition of new nodes and links and removal of some existing nodes and links in *G*.

*Model +N:* Figure 2(b) represents the updated network $G_0$. During the update, two new nodes *V11* and *V12* are introduced. $G_0$ is a disconnected network containing three components. It contains *n= 12* and *m=11*.

*Model -N:* Figure 2(c) indicates a new network $G_1$. During the change, two nodes *V3* and *V4* are removed from *G*. $G_1$ contains *n = 8* and *m =9*.

*Model ±N:* Figure 2(d) represents the updated network $G_2$. In *G*, *V11* and *V12* are added and *V3* and *V4* are removed. $G_2$ contains *n = 10* and *m = 9*.

*Model +L:* Figure 2(e) depicts the snapshot of the graph $G_3$ after addition of *V1-V2* and *V5-V9*. It contains *n=10* and *m=14*.

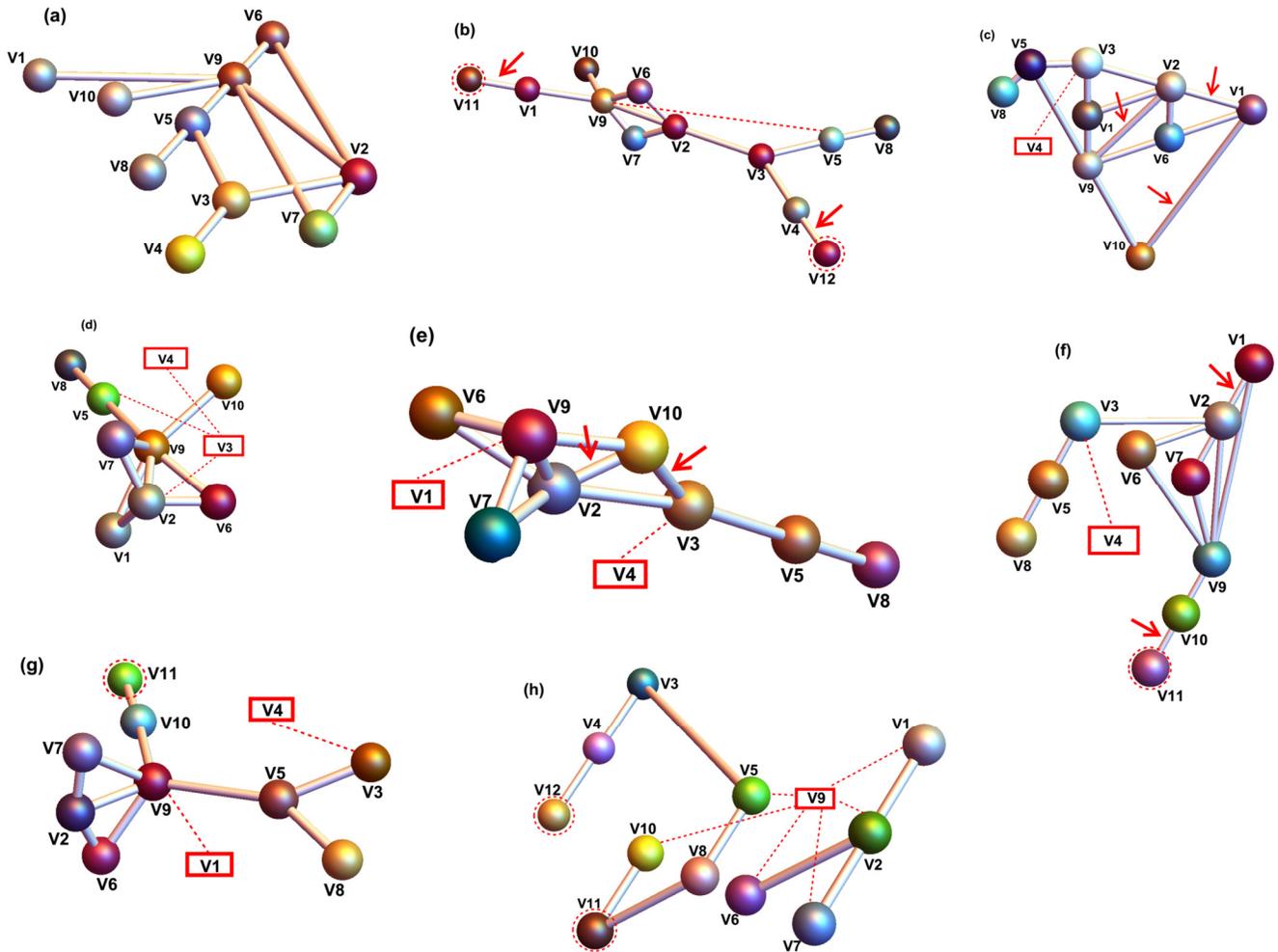

**Figure 3.** The dotted circles and rectangle nodes represent added and deleted nodes respectively. Red arrows and dotted lines represent the addition and deletion of links in the network respectively. a) The random network *G*, containing *N=10* and *L=12;* b) The network obtained after addition of new nodes as well as links and removal of links in *G*; c) Network obtained after removal of nodes and addition of new links in *G*; d) Network obtained after removal of existing links and nodes in *G*; e) The network obtained after addition of new links as well as the new nodes and removal of existing links in *G*; f) Network obtained after addition of new nodes as well as the links and removal of some existing nodes in *G*; g) Network obtained after addition of new nodes and removal of some existing nodes as well as links in *G*; h) Network obtained after addition of new nodes and links and removal of some existing nodes and links in *G*.

*Model -L:* Figure 2(f) represents the graph $G_4$, obtained by the removal of *V9-V10, V2-V9*, and *V5-V9* from *G*. $G_4$ contains *n=10* and *m=9*.

*Model ±L:* Figure 2(g) represents the graph $G_5$, obtained by adding *V3-V10* and *V8-V9* and removing *V5-V9* and *V5-V6* in *G*. $G_5$ contains *n=10* and *m=12*.

*Model +N+L:* Figure 2(h) displays graph $G_6$, which demonstrates the change in *G* by the addition of *V11, V12, V1-V11, V8-V12,* and *V10-V12*. $G_6$ contains *n=12* and *m=14*.

*Model +N-L:* Figure 2(i) illustrates graph $G_7$ obtained by the addition of *V11* and removal of *V3-V4* and *V5-V8* in *G*. $G_7$ contains *n=11* and *m=9*.

*Model +N±L:* Figure 3(b) depicts the graph $G_8$ obtained by the addition of *V11, V12, V1-V11* and *V4-V12* and removal of *V5-V9* in *G*. $G_8$ contains *n=12 and m=13*.

*Model -N+L:* Figure 3(c) represents graph $G_9$, which is a change in the network *G* by removal of *V4* and addition of *V1-V2, V1-V10,* and *V3-V9*. $G_9$ contains *n=9* and *m=14*.

*Model -N-L:* Figure 3(d) demonstrates graph $G_{10}$, obtained by removal of *V3, V4, V2-V3, V3-V5,* and *V3-V4* in *G*. $G_{10}$ contains *n=8* and *m=10*.

*Model -N±L:* Figure 3(e) illustrates graph $G_{11}$, a snapshot of *G* after removal of *V1, V4, V3-V4* and *V1-V9* and the addition of *V2-V10* and *V3-V10*. $G_{11}$ contains *n=8* and *m=11*.

*Model ±N+L:* Figure 3(f) displays graph $G_{12}$, an update of the network *G* by the removal of *V4* and addition of *V11, V1-V2,* and *V10-V11*. $G_{12}$ contains *n=10* and *m=12*.

*Model ±N-L:* Figure 3(g) depicts the network $G_{13}$ after an update in the network *G* by removal of *V1, V4,* and *V1-V9* and addition of *V11*. $G_{13}$ contains *n=9* and *m=10*.

*Model ±N±L:* Figure 3(h) illustrates the network $G_{14}$, that is an update of *G* obtained by the addition of *V11*, *V12*, *V4-V12*, *V10-V11*, and *V8-V11* and removal of *V9, V1-V9, V2-V9, V5-V9, V6-V9, V7-V9,* and *V9-V10*. $G_{14}$ contains *n=11* and *m=9*.

### III. WHERE TO GO FROM HERE?

#### A. Complex Network Tools: An Overview and a Short Survey

An effective set of mathematical, statistical, and computational tools for complex network analysis and modeling is available. Every tool has a different architecture and computational models; one size does not fit all. Some are specialized, some are generic; some are single threaded, some or multi-threaded, some have shared memory, some have distributed memory. Network scientists use these tools for network construction, visualization, simulation, properties extraction, pattern recognition, and complex queries.

Some of the commonly used network analysis and visualization tools include *Pajek, Wolfram Mathematica, MATLAB, R, Gephi, Visone,* and *CiteSpace*.

#### B. Review of Network Books

There is a wide variety of books to model and analyze complex networks, such as [6] describes basics of networks, [8, 10] presents the statistical analysis of networks, [4, 14] provides social network analysis, [17] demonstrates structure and dynamics of complex networks.

#### C. Discussion and Conclusions

Communication engineers and researchers often need to work on systems involving a large number of interacting entities. In the past, these nodes were often limited both in complexity as well as in number.

Recent advances in networks such as Body Sensor Networks (BSN), Wireless Sensor Networks (WSN), and Internet of Things (IoT) as well as Vehicular Networks (VANETS) has resulted in a proliferation of interconnected electronic communication devices. As such, the requirements of handling, understanding and modeling large sets of interacting nodes is no longer a novelty but has rather become a much-needed tool of the trade. It is therefore an essential skill for addition to the repertoire of any communications system professional/researcher.

Quite simply, every network designer now needs the ability to understand the important characteristics of their system as it grows in number as well as complexity. In the absence of modeling and simulation exercises, there are bound to be problems in the resultant network. While tools for dissecting complexity in large-scale systems with numerous interacting entities have evolved over time for interdisciplinary researchers such as from the social sciences or physics.

The interaction in these modern communication networks could be mapped as a graph. For example, in IoT, smart devices could be the nodes and internet/wireless connections could be the links; in WSN, sensor devices form individual nodes and wireless medium form links; in BSN, the implementable and wearable sensors represent the nodes with the radio signals being modeled as links; and in physical networks, such as LAN contains machines as nodes and wireless or communication lines as links.

Traditional methods of modeling are only able to capture certain limited aspects of a network. They are unable to capture such complex network dynamics and emergent behavior. Representation of communication networks in this form allows for treating the communication network as a complex network. This eventually allows for conducting a topological analysis. While complex networks provide a formal framework for modeling communication systems, this knowledge is not very well known in the communication systems' community.

In this article, we present a comprehensive overview of complex communication networks modeling and analysis using techniques from graphs and social/complex network analysis. We also examine how network dynamics can be represented using complex networks. We have used a complex network approach to analyze these networks in more efficient and effective way to identify important nodes, influential nodes, the nodes creating bottleneck or congestion, bridges, network centrality, individual node's prominence with a network, community structure, and the broker nodes. Our goal was to teach communication engineers and researchers how to understand the dynamics of complex communication networks modeled as graphs. We have also given a comprehensive overview of the common network analysis techniques. In future, this work can be extended to simulate dynamics on and of networks.

## References

[1]     F. T. Boesch, F. Harary, and J. A. Kabell, "Graphs as models of communication network vulnerability: Connectivity and

persistence," *Networks,* vol. 11, no. 1, pp. 57-63, 1981.
[2] H. Sayama, I. Pestov, J. Schmidt, B. J. Bush, C. Wong, J. Yamanoi, and T. Gross, "Modeling complex systems with adaptive networks," *Computers & Mathematics with Applications*, 2013.
[3] S. Fortunato, and D. Hric, "Community detection in networks: A user guide," *Physics Reports,* vol. 659, pp. 1-44, 2016.
[4] W. De Nooy, A. Mrvar, and V. Batagelj, *Exploratory social network analysis with Pajek*: Cambridge University Press, 2011.
[5] T. V. Cavalcanti, C. Giannitsarou, and C. R. Johnson, "Network cohesion," *Economic Theory*, pp. 1-21, 2012.
[6] M. Newman, "Networks: an introduction. 2010," *United Slates: Oxford University Press Inc., New York*, pp. 1-2.
[7] P. Erdos, and A. Rényi, "On the evolution of random graphs," *Publ. Math. Inst. Hung. Acad. Sci,* vol. 5, no. 1, pp. 17-60, 1960.
[8] R. Albert, and A.-L. Barabási, "Statistical mechanics of complex networks," *Reviews of modern physics,* vol. 74, no. 1, pp. 47, 2002.
[9] D. J. Watts, and S. H. Strogatz, "Collective dynamics of 'small-world' networks," *nature,* vol. 393, no. 6684, pp. 440-442, 1998.
[10] D. A. Luke, *A user's guide to network analysis in R*: Springer, 2016.
[11] K. Batool, and M. A. Niazi, "Towards a methodology for validation of centrality measures in complex networks," *PloS one,* vol. 9, no. 4, pp. e90283, 2014.
[12] M. Dehmer, and F. Emmert-Streib, *Analysis of complex networks: from biology to linguistics*: John Wiley & Sons, 2009.
[13] S. Boccaletti, V. Latora, Y. Moreno, M. Chavez, and D.-U. Hwang, "Complex networks: Structure and dynamics," *Physics reports,* vol. 424, no. 4, pp. 175-308, 2006.
[14] U. Brandes, and D. Wagner, "Analysis and visualization of social networks," *Graph drawing software*, pp. 321-340: Springer, 2004.
[15] M. A. N. Komal Batool, "Modeling the Internet of Things: a hybrid modeling approach using complex networks and agent based modeling," *Complex adaptive system modeling,* vol. 5:4, 2017.
[16] M. A. Niazi, and A. Hussain, *Cognitive agent-based computing-I: a unified framework for modeling complex adaptive systems using agent-based & complex network-based methods*: Springer Science & Business Media, 2012.
[17] M. Newman, A.-L. Barabasi, and D. J. Watts, *The structure and dynamics of networks*: Princeton University Press, 2011.

BIOGRAPHIES

SHAMAILA BISMA KHAN (bis.sarfraz@gmail.com) is a PhD student at COMSATS Institute of IT, Islamabad, Pakistan. Her research interest includes the formal methods, citation networks, and complex dynamic networks.

MUAZ AHMAD NIAZI (muaz.niazi@ieee.org) is a Chief Scientific Officer (Professor) at COMSATS Institute of IT, Islamabad, Pakistan. With an undergraduate degree in Electrical Engineering, he has an MS and a PhD in Computer Sciences from Boston University, MA, USA and the University of Stirling, Scotland, UK respectively in addition to a postdoc from the University of Stirling's COSIPRA Lab.